\documentclass[]{article}
\makeatletter
\usepackage{paralist}
\usepackage{enumitem}
\usepackage{booktabs}
\usepackage{url}
\usepackage{amsmath}
\usepackage{amsthm}

\newtheorem{theorem}{Theorem}

\usepackage[letterpaper, total={6in, 8in}]{geometry}

\@ifundefined{showcaptionsetup}{}{%
 \PassOptionsToPackage{caption=false}{subfig}}
\usepackage{subfig}

\PassOptionsToPackage{normalem}{ulem}
\usepackage{ulem}

\providecommand{\tabularnewline}{\\}

\usepackage{tikz}
\usetikzlibrary{bayesnet}

\makeatletter

\title{On the Equivalence of Generative and Discriminative Formulations
of the Sequential Dependence Model}

\author{
  Laura Dietz\\
  University of New Hampshire\\
  \texttt{dietz@cs.unh.edu}
  \and
  John Foley\\
  University of Massachusetts \\
  \texttt{jfoley@cs.umass.edu}
}

\date{}

\begin{document}

\maketitle

\begin{abstract}

The sequential dependence model (SDM) is a popular retrieval model which
is based on the theory of probabilistic graphical models. While it
was originally introduced by Metzler and Croft as a Markov Random
Field (aka discriminative probabilistic model), in this paper we demonstrate that it is equivalent to a generative probabilistic model. 

To build an foundation for future retrieval models, this paper details the axiomatic underpinning of the SDM model as discriminative and generative probabilistic model. The only difference arises whether model parameters are estimated in log-space or Multinomial-space. We demonstrate that parameter-estimation with grid-tuning is negatively impacting the generative formulation, an effect that vanishes when parameters are estimated with coordinate-gradient descent. This is concerning, since empirical differences may be falsely attributed to improved models.\footnote{This paper was also presented at the SIGIR'17 Workshop on Axiomatic Thinking for Information Retrieval and Related Tasks (ATIR).}

%
\end{abstract}

\section{Introduction}

The sequential dependence model \cite{metzler2005sdm} is a very robust
retrieval model that has been shown to outperform or to be on par
with many retrieval models \cite{huston2013termdependencies}. Its
robustness comes from an integration of unigram, bigram, and windowed
bigram models through the theoretical framework of Markov random fields.
The SDM Markov random field is associated with a set of parameters
which are learned through the usual parameter estimation techniques
for undirected graphical models with training data. Despite its simplicity,
the SDM model is a versatile method that provides a reasonable input
ranking for further learning-to-rank phases or in as a building block
in a larger model \cite{dalton2013kbbridge}. As it is a feature-based
learning-to-rank model, it can be extended with additional features,
such as in the latent concept model \cite{bendersky2011quality,metzler2007lce}.
Like all Markov random field models it can be extended with further
variables, for instance to incorporate external knowledge, such as
entities from an external semantic network. It can also be extended
with additional conditional dependencies, such as further term dependencies
that are expected to be helpful for the retrieval task, such as in
the hypergraph retrieval model \cite{bendersky2012hypergraphs}. 

The essential idea of the sequential dependence model (SDM) is to
combine unigram, bigram, and windowed bigram models so that they mutually
compensate each other's shortcomings. The unigram gram model, which
is also called the bag-of-words model and which is closely related
to the vector-space model, is indifferent to word order. This is an
issue for multi-word expressions which are for instance common for
entity names such as ``Massachusetts Institute of Technology'' or
compound nouns such as ``information retrieval'' which have a different
meaning in combination than individually. This shortcoming is compensated
for in bigram model which incorporate word-order by modeling the probability
of joint occurrence of two subsequent query words $q_{i-1}q_{i}$
or condition the probability of $i$th word in the query, $q_{i}$,
on seeing the previous word $q_{i-1}$. 

One additional concern is that users tend to remove non-essential
words from the information need when formulating the query, such as
in the example query ``prevent rain basement'' to represent the
query ``how can I prevent the heavy spring rain from leaking into
my brick house's basement?''. The bigram model which only captures
consecutive words may not be able to address this situation. This
motivates the use of bigram models that allow for length-restricted
gaps. Literature describes different variants such models under the
names skip gram models or orthogonal sparse bigrams \cite{siefkes2004osb}.
In this work, we focus on a variant that has been used successfully
in the sequential dependence model, which models the co-occurrence
of two terms within a window of eight\footnote{The window length requires tuning in practice; we follow the choice
of eight for compliance with previous work.} terms, which we refer to as windowed bigrams.

The sequential dependence model combines ideas of all three models
in order to compensate respective shortcomings. The retrieval model
scores documents for a query though the theoretical framework of Markov
random field models (MRF). However, there are a set of related models
that address the same task and originate from generative models and
Jelinek-Mercer smoothing. In addition, different variants of bigram
models have been used interchangeably, i.e., based on a bag-of-bigrams
approach and an n-gram model approach which leads to different scoring
algorithms. A decade after the seminal work on the sequential dependence
model has been published, we aim to reconsider some of the derivations,
approximations, and study similarities and differences arising from
several choices. Where Huston et al.\ \cite{huston2013termdependencies,huston2014termdepedencies-appendix}
emphasized a strictly empirical study, in this work we reconsider
the SDM model from a theoretical side. The contributions of this paper
are the following.
\begin{itemize}
\item Theoretical analysis of similarities and differences for MRF versus
other modelling frameworks and different bigram paradigms.
\item Empirical study on effects on the retrieval performance and weight
parameters estimated\footnote{Code and runs available: \url{https://bitbucket.org/jfoley/prob-sdm}}.
\item Discussion of approximations made in an available SDM implementation
in the open-source engine Galago.
\end{itemize}

\paragraph{Outline}

After clarifying the notation, we state in Section \ref{sec:Sequential-Dependence-Algo}
the SDM scoring algorithm with Dirichlet smoothing as implemented
in the search engine Galago V3.7. In Section \ref{sec:mrfSDM} we
recap the original derivation of this algorithm as a Markov Random
Field. A generative alternative is discussed in Section \ref{sec:genSDM}
with connections to MRF and Jelinek-Mercer models. Where this is modeling
bigrams with the bag-of-bigrams approach, Section \ref{sec:genNgram}
elaborates on an alternative model that is following the n-gram model
approach instead. Section \ref{sec:Experiments} demonstrates the empirical equivalence the different
models when proper parameter learning methods are used. Related work is discussed
in Section \ref{sec:Related-work} before we conclude.

\section{Notation}

\label{sec:Notation}We refer to a model as $\mathcal{M}$, and the
likelihood of data under the model as $\mathcal{L}_{\mathcal{M}}$,
and a probability distribution of a variable as $p(X)$. We refer
to the numerical ranking score function provided by the model $\mathcal{M}$ for given arguments
as $\mbox{score}_{\mathcal{M}}(...)$. For graphical models, this
score is rank-equivalent to the model likelihood; or equivalently the log-likelihood.
In correspondence to conditional probabilities $\mathcal{L}(X|Y)$ we
refer to rank equivalent expressions to conditional scores, i.e.,
$\mbox{score}(X|Y)$.

We refer to counts as $n$ with subscripts. For example, a number
of occurrences of a term $w$ in a document $d$ is denoted $n_{w,d}$.
To avoid clutter for marginal counts, i.e., when summing over all counts
for possible variable settings, we refer to marginal counts as $\star$.
For example, $n_{\star,d}$ refers to all words in the document (also
sometimes denoted as $|d|$), while $n_{w,\star}$ refers to all occurrences
of the word $w$ in any document, which is sometimes denoted as $cf(w)$.
Finally, $n_{\star,\star}=|C|$ denotes the total collection frequency.
The vocabulary over all terms is denoted $V$.

We distinguish between random variables by uppercase notation, e.g.
$Q$, $D$, and concrete configurations that the random variables
can take on, as lower case, e.g., $q$, $d$. Feature functions of
variable settings $x$ and $y$ are denoted as $\mbox{\ensuremath{\mathbf{f}}}(x,y)$.
We denote distribution parameters and weight parameters as greek letters.
Vector-valued variables are indicated through bold symbols, e.g., $\boldsymbol{\lambda}$,
while elements of the vector are indicated with a subscript, e.g.
$\lambda_{u}$.

\section{Sequential Dependence Scoring Implementation}

\label{sec:Sequential-Dependence-Algo}Given a query $\mathbf{q}=q_{1},q_{2},\dots,q_{k}$,
the sequential dependence scoring algorithm assigns a rank-score for
each document $d$. The algorithm further needs to be given as parameters
$\boldsymbol{\lambda}=\lambda_{u},\lambda_{b},\lambda_{w}$ which
are the relative weights trading-off unigram (u), bigram (b), and
windowed-bigram (w) models. 

Using shorthand $\mathcal{M}_{u}$ for the unigram language model,
$\mathcal{M}_{b}$ for the bigram language model, and $\mathcal{M}_{w}$
for an unordered-window-8 language model, the SDM score for the document
$d$ is computed as,

\begin{eqnarray}
\mbox{score}_{SDM}(d|\mathbf{q},\boldsymbol{\lambda}) & = & \lambda_{u}\mbox{\ensuremath{\cdot}}\mbox{score}_{\mathcal{M}_{u}}(d|\mathbf{q})+\lambda_{b}\mbox{\ensuremath{\cdot}}\mbox{score}_{\mathcal{M}_{b}}(d|\mathbf{q}) + \lambda_{w}\mbox{\ensuremath{\cdot}}\mbox{score}_{\mathcal{M}_{w}}(d|\mathbf{q})\label{eq:sdm-scoring}
\end{eqnarray}

While the algorithm is indifferent towards the exact language models
used, the common choice is to use language models with smoothing. The
original work on SDM uses Jelinek-Mercer smoothing. Here, we first
focus on Dirichlet smoothing to elaborate on connections to generative
approaches. Dirichlet smoothing requires an additional parameter $\mu$
to control the smoothing trade-off between the document and the collection
statistics.

\paragraph*{Unigram model}

$\mathcal{M}_{u}$ also refers to the query likelihood model, which is
represented by the inquery \cite{callan1995inquery} operator \texttt{\#combine(}$q_{1}$
$q_{2}$ $\dots$ $q_{k}$\texttt{)}. Using Dirichlet smoothing, this operator implements the following
scoring equation.

\begin{equation}
\mbox{score}_{\mathcal{M}_{u}}(d|\mathbf{q})=\sum_{q_{i}\in\mathbf{q}}\log\frac{n_{q_{i},d}+\mu\frac{n_{q_{i},\star}}{n_{\star,\star}}}{n_{\star,d}+\mu}\label{eq:unigram-model}
\end{equation}

where, $n_{\star,d}$ is the document length, and $n_{\star,\star}$
denotes the number of tokens in the corpus. To underline the origin
of sums, we use the notation for sums over all elements in a vector, e.g.
$\sum_{q_{i}\in\mathbf{q}}\dots$ for all query terms, instead of
the equivalent notation of sums over a range indices of the vector,
e.g., $\sum_{i=1}^{k}\dots$.

\paragraph*{Bigram model}

For $\mathcal{M}_{b}$, a common choice is an ordered bigram model
with Dirichlet smoothing, which is represented by the inquery operator
chain \texttt{\#combine(\#ordered:1($q_{1}$ $q_{2}$) \#ordered:1( $q_{2}$
$q_{3}$) $\ldots$ \#ordered:1($q_{k-1}$ $q_{k}$))}. With Dirichlet
smoothing, this operator-chain implements the scoring function,

\[
\mbox{score}_{\mathcal{M}_{b}}(d|\mathbf{q})=\sum_{\left(q_{i},q_{i+1}\right)\in\mathbf{q}}\log\frac{n_{\left(q_{i},q_{i+1}\right),d}+\mu\frac{n_{\left(q_{i},q_{i+1}\right),\star}}{n_{\left(\star,\star\right),\star}}}{n_{\left(\star,\star\right),d}+\mu}
\]

where, $n_{\left(q_{i},q_{i+1}\right),d}$ denotes the number of bigrams $q_{i} \circ q_{i+1}$
occurring in the document. The number of bigrams in the document,
$n_{\left(\star,\star\right),d}=|d|-1$, equals the document length
minus one.

\paragraph*{Windowed-Bigram model}

For the windowed\nobreakdash-bigram model $\mathcal{M}_{w}$, a common choice
is to use a window of eight terms and ignoring the word order. Note
that word order is only relaxed on the document side, but not on the
query side, therefore only consecutive query terms $q_{i}$ and $q_{i+1}$
are considered. This is represented by the inquery operator chain
\texttt{\#combine(\#unordered:8($q_{1}$ $q_{2}$) \#unordered:8($q_{2}$
$q_{3}$) \textcompwordmark{} $\ldots$ \#unordered:8($q_{k-1}$
$q_{k}$))}. With Dirichlet smoothing of empirical distributions over
windowed bigrams, this operator-chain implements the scoring function,

\[
\mbox{score}_{\mathcal{M}_{w}}(d|\mathbf{q})=\sum_{\left(q_{i},q_{i+1}\right)\in\mathbf{q}}\log\frac{n_{\left\{ q_{i},q_{i+1}\right\} _{8},d}+\mu\frac{n_{\left\{ q_{i},q_{i+1}\right\} _{8},\star}}{n_{\left\{ \star,\star\right\} _{8},\star}}}{n_{\left\{ \star,\star\right\} _{8},d}+\mu}
\]

where $n_{\left\{ q_{i},q_{i+1}\right\} _{8},d}$ refers to the number
of times the query terms $q_{i}$ and $q_{i+1}$ occur within eight
terms of each other. 

\paragraph*{Implementation-specific approximations}
The implementation within Galgo makes several approximations on collection counts for bigrams as $n_{\left\{ \star,\star\right\} _{8},d} \approx n_{\left(\star,\star\right),d} \approx n_{\star,d}=|d|$. This approximation is reasonable in some cases, as we discuss in the appendix.

\section{Markov Random Field Sequential Dependence Model}

\label{sec:mrfSDM}In this Section we recap the derivation of the
SDM scoring algorithm.

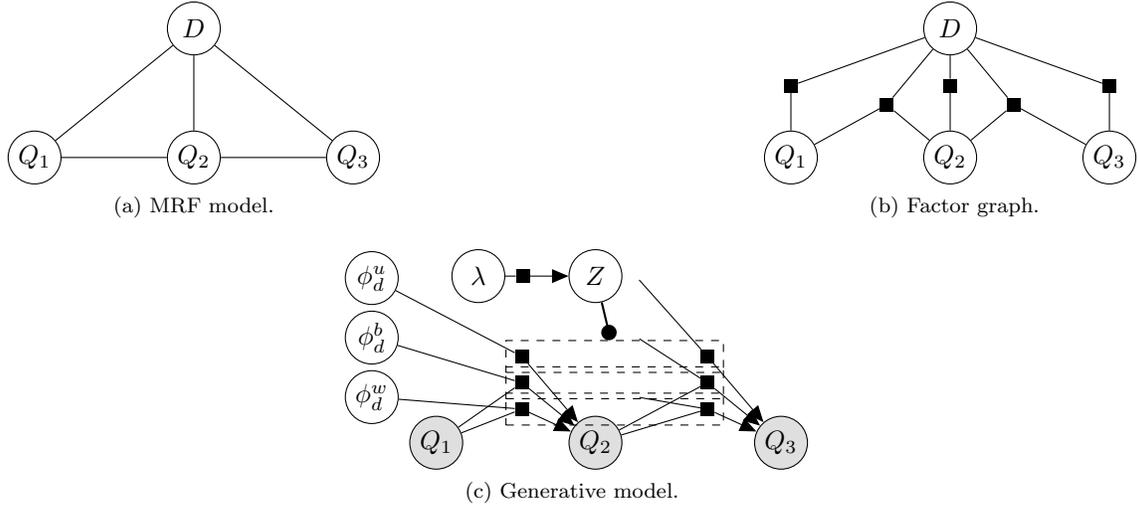
\begin{figure*}
\subfloat[MRF model.\label{fig:MRF-model}]{
\begin{tikzpicture}[x=0.7cm,y=0.5cm]

  \node[latent]                   (D)      {$D$} ; %
  \node[latent, below=2 of D]    (Q2)      {$Q_2$} ; %
\node[latent, left=2 of Q2]    (Q1)      {$Q_1$} ; %
\node[latent, right=2 of Q2]    (Q3)      {$Q_3$} ; %


\factoredge {D}  {Q1}  {} ; %
\factoredge {D}  {Q2}  {} ; %
\factoredge {D}  {Q3}  {} ; %
\factoredge {Q1}  {Q2}  {} ; %
\factoredge {Q2}  {Q3}  {} ; %


\end{tikzpicture}

} \hfill{}\subfloat[Factor graph.\label{fig:Factor-graph}]{\begin{tikzpicture}[x=0.7cm,y=0.5cm]

  \node[latent]                   (D)      {$D$} ; %
  \node[latent, below=2 of D]    (Q2)      {$Q_2$} ; %
\node[latent, left=2 of Q2]    (Q1)      {$Q_1$} ; %
\node[latent, right=2 of Q2]    (Q3)      {$Q_3$} ; %

\factor[above left=1 of Q2]     {q1q2d}     {} {} {} ; %
\factor[above right=1 of Q2]     {q2q3d}     {} {} {} ; %

\factor[above=1 of Q1]     {q1d}     {} {} {} ; %
\factor[above=1 of Q2]     {q2d}     {} {} {} ; %
\factor[above=1 of Q3]     {q3d}     {} {} {} ; %


\draw (D) -- (q1q2d) ; %
\draw (Q1) -- (q1q2d) ; %
\draw (Q2) -- (q1q2d) ; %

\draw (D) -- (q2q3d) ; %
\draw (Q2) -- (q2q3d) ; %
\draw (Q3) -- (q2q3d) ; %

\draw (D) -- (q1d) -- (Q1) ; %
\draw (D) -- (q2d) -- (Q2) ; %
\draw (D) -- (q3d) -- (Q3) ; %


\end{tikzpicture}
}

\centering{ \subfloat[Generative model.\label{fig:Generative-model}]{
\begin{tikzpicture}[x=0.7cm,y=0.5cm]


\node[latent] (phiu)	{$\phi^u_d$} ; %
\node[latent, below=0.15 of phiu] (phib)	{$\phi^b_d$} ; %
\node[latent, below=0.15 of phib] (phiw)	{$\phi^w_d$} ; %

\node[obs, below right=0.2 and 0.5 of phiw]    (Q1)      {$Q_1$} ; %
\node[obs, right=2 of Q1]    (Q2)      {$Q_2$} ; %
\node[obs, right=2.5 of Q2]    (Q3)      {$Q_3$} ; %

\node[latent, above=3 of Q2]    (Z2)      {$Z$} ; %

\node[latent, left=1.2 of Z2]	(lambda)	{$\lambda$} ; %

\node[const, above left=-0.3 and 1 of Q2] (marker) {} ; %

\factor[above=1.8 of marker]     {q2u}     {} {} {} ; %
\factor[above=1.1 of marker]     {q2b}     {} {} {} ; %
\factor[above=0.4 of marker]     {q2w}     {} {} {} ; %

\factoredge {phiu}  {q2u}  {Q2} ; %
\factoredge {phib}  {q2b}  {Q2} ; %
\factoredge {phiw}  {q2w}  {Q2} ; %

\draw (Q1) -- (q2b) ; %
\draw (Q1) -- (q2w) ; %

\node[const, above left=-0.3 and 1 of Q3] (marker2) {} ; %

\factor[above=1.8 of marker2]     {q3u}     {} {} {} ; %
\factor[above=1.1 of marker2]     {q3b}     {} {} {} ; %
\factor[above=0.4 of marker2]     {q3w}     {} {} {} ; %
\factor[right=0.2 of lambda]	{lambda2}	{} {} {} ;%

\node[const, right=4.5 of phiu] (phiu2) {}; %
\node[const, right=4.5 of phib] (phib2) {}; %
\node[const, right=4.5 of phiw] (phiw2) {}; %

\factoredge {phiu2}  {q3u}  {Q3} ; %
\factoredge {phib2}  {q3b}  {Q3} ; %
\factoredge {phiw2}  {q3w}  {Q3} ; %

\draw (Q2) -- (q3b) ; %
\draw (Q2) -- (q3w) ; %

\factoredge {lambda}	{lambda2} 	{Z2} ; %

\gate  {z2gate1} {(q2u) (q3u)} {Z2} ; %
\gate  {z2gate2} {(q2b) (q3b)} {} ; %
\gate  {z2gate3} {(q2w) (q3w)} {} ; %

\end{tikzpicture}}
}

\caption{Sequential Dependence Model.}

\end{figure*}

Metzler et al. derive the algorithm in Section 3 through a Markov
Random Field model for term dependencies, which we recap in this section.
Markov random fields, which are also called undirected graphical models,
provide a probabilistic framework for inference of random variables
and parameter learning. A graphical model is defined to be a Markov
random field if the distribution of a random variable only depends
on the knowledge of the outcome of neighboring variables. We limit
the introduction of MRFs to concepts that are required to follow the
derivation of the Sequential Dependence Model, for a complete introduction
we refer the reader to Chapter 19.3 of the text book of Murphy \cite{murphy2012book}. 

To model a query $\mathbf{q}=q_{1}q_{2}\ldots q_{k}$ and a document
$d$, Metzler et al.\ introduce a random variable $Q_{i}$ for each
query term $q_{i}$ as well as the random variable $D$ to denote
a document $d$ from the corpus which is to be scored. For example, $Q_{1}=\mbox{'information'}$, $Q_{2}=\mbox{'retrieval'}$.  The sequential
dependence model captures statistical dependence between random variables of consecutive
query terms $Q_{i}$ and $Q_{i+1}$ and the document $D$, cf.\ Figure
\ref{fig:MRF-model}. 

However, non-consecutive query terms $Q_{i}$ and $Q_{j}$ (called
non-neighbors) are intended to be conditionally independent, given
the terms in between. By rules of the MRF framework, unconnected random
variables are conditionally independent given values of remaining
random variables. Therefore, the absence of connections between non-neighbors
$Q_{i}$ and $Q_{j}$ in the Graphical model (Figure \ref{fig:MRF-model})
declares this independence.

The framework of Markov Random Fields allows to reason about observed
variables and latent variables. As a special case of MRFs, all variables of the
sequential dependence model are observed. This means that we know
the configuration of all variables during inference relieving us from
treating unknowns. The purpose of MRFs for the sequential dependence
scoring algorithm is to use the model likelihood $\mathcal{L}$ as
a ranking score function for a document $d$ given the query terms
$\mathbf{q}$.

\subsection{SDM Model Likelihood}

The likelihood $\mathcal{L}$ of the sequential dependence model for
a given configuration of the random variables $Q_{i}=q_{i}$ and $D=d$
provides the retrieval score for the document $d$ given the query
$\mathbf{q}$.

According to the Hammersley-Clifford theorem \cite{murphy2012book},
the likelihood $\mathcal{L}$ (or joint distribution) of a Markov
Random Field can be fully expressed over a product over maximal cliques
in the model, where each clique of random variables is associated
with a nonnegative potential function $\psi$. For instance in the
sequential dependence model, a potential function $\psi$ for the
random variables $Q_{1},Q_{2},$ and $D$, produces a nonnegative
real-valued number for every configuration of the random variables
such as $Q_{1}=\mbox{'information'}$, $Q_{2}=\mbox{'retrieval'}$,
and $D$ referring to a document in the collection.

The Hammersley-Clifford theorem states that it is possible to express
the likelihood of every MRF through a product over maximal cliques
(not requiring further factors over unconnected variables). However,
the theorem does not provide a constructive recipe to do so. Instead,
it is part of devising the model to choose a factorization of the
likelihood into arbitrary cliques of random variables. Where the MRF
notation only informs on conditional \uline{\emph{in}}dependence, the equivalent
graphical notation of factor graphs additionally specifies the factorization
chosen for the model, cf.\ Figure \ref{fig:Factor-graph}. 

In the factor graph formalization, any set of variables that form
a factor in the likelihood are connected to a small box. A consistent
factor graph of the sequential dependence model is given in Figure
\ref{fig:Factor-graph}. The equivalent model likelihood for the sequential
dependence model follows as,

\[
\mathcal{L}(\mathbf{Q},D)=\frac{1}{Z(\boldsymbol{\lambda})}\prod_{q_{i}\in\mathbf{q}}\psi(Q_{i},D|\boldsymbol{\lambda})\cdot\prod_{Q_{i},Q_{i+1}\in\mathbf{Q}}\psi(Q_{i},Q_{i+1},D|\boldsymbol{\lambda})
\]

\noindent Where $Z(\boldsymbol{\lambda})$, the partition
function, is a constant that ensures normalization
of the joint distribution over all possible configurations of $Q_{i}\in V$
and all documents $d$. This means that summing $\mathcal{L}$ over
all possible combinations of query terms in the vocabulary $V$ and
all documents in the corpus will sum to 1. 

However, as the sequential dependence model is only used to rank documents
for a given query $\mathbf{q}$ by the model likelihood $\mathcal{L}$,
the constant $Z(\boldsymbol{\lambda})$ can be ignored to provide a rank
equivalent scoring criterion $\mbox{score}_{SDM}$. 

\subsection{Ranking Scoring Criterion}

With the goal of ranking elements by the SDM likelihood function,
we can alternatively use any other rank-equivalent criterion. For instance,
we equivalently use the log-likelihood $\log\mathcal{L}$ for scoring,
leaving us with the following scoring criterion.
\begin{align}
 \mbox{score}_{SDM}(d|\mathbf{q})& \; \stackrel{rank}{=} \; \log\mathcal{L}(\mathbf{q},d)\label{eq:sdm-factorized}\\
 & \; \stackrel{rank}{=} \; \sum_{q_{i}\in\mathbf{q}}\log\psi(Q_{i},D|\boldsymbol{\lambda})+\sum_{q_{i},q_{i+1}\in\mathbf{q}}\log\psi(Q_{i},Q_{i+1},D|\boldsymbol{\lambda})\nonumber
\end{align}

%
%

\paragraph{Potential functions}

The MRF framework provides us with the freedom to choose the functional
form of potential functions $\psi$. The only hard restriction implied
by MRFs is that potential functions ought to be nonnegative. When
considering potential functions in log-space, this means that the
quantity $\log\psi$ can take on any real value while being defined
on all inputs. 

The sequential dependence model follows a common choice by using a
so-called log-linear model as the functional form of the potentials
$\log\psi$. The log-linear model is defined as an inner product of
a feature vector $\mbox{\ensuremath{\mathbf{f}}}(\dots)$ and a parameter
vector $\boldsymbol{\lambda}$ in log-space. The entries of the feature
vector are induced by configurations of random variables in the clique
which should represent a measure of compatibility between different
variable configurations.

For instance in the sequential dependence model, the clique of random
variables $Q_{1}$, $Q_{2},$ and $D$ is represented as a feature
vector of a particular configuration $Q_{1}=q_{1}$, $Q_{2}=q_{2}$,
and $D=d$ which is denoted as $\mbox{\ensuremath{\mathbf{f}}}(q_{1},q_{2},d)$.
The log-potential function is defined as the inner product between
the feature vector and a parameter vector $\boldsymbol{\lambda}$
as

\begin{eqnarray*}
\log\psi(Q_{1},Q_{2},D|\boldsymbol{\lambda}) & = & \sum_{j=1}^{m}\mbox{f}_{j}(q_{1},q_{2},d)\cdot\lambda_{j}
\end{eqnarray*}

where $m$ denotes the length of the feature vector
or the parameter vector respectively. Each entry of the feature vector,
$\mbox{f}_{j}$ should express compatibility of the given variable
configurations, to which the corresponding entry in the parameter
vector $\lambda_{j}$ assigns relative weight. Since we operate in
log-space, both positive and negative weights are acceptable.

\paragraph{Factors and features}

The sequential dependence model makes use of two factor types, one
for the two-cliques of for single query terms and the document, and
another for the three-cliques of consecutive query terms and the document.
Both factor types are repeated across all query terms. Each factor
type goes along with its own feature vector functions and corresponding
parameter vector. While not necessarily the case, in this model, the
same parameter vector is shared between all factors of the same factor
type (so-called parameter-tying).

The sequential dependence model associates each two-clique $\log\psi(Q_{i},D|\boldsymbol{\lambda});\forall i$
with a feature vector of length one, consisting only of the unigram
score of $q_{i}$ in the document $d$, denoted by Equation \ref{fu}.
The three-clique $\log\psi(Q_{i-1},Q_{i},D|\boldsymbol{\lambda});$
$\forall i\geq2$ is associated with a feature vector of length two,
consisting of the bigram score of $q_{i-1}$ and $q_{i}$ in the document,
denoted Equation \ref{fb}, as well as the windowed-bigram score Equation
\ref{fw}. 

\begin{eqnarray}
\mbox{f}_{u}(q_{i},d) & = & \mbox{score}_{\mathcal{M}_{u}}(d|q_{i})\label{fu}\\
\mbox{f}_{b}(q_{i-1},q_{i},d) & = & \mbox{score}_{\mathcal{M}_{b}}(d|q_{i-1},q_{i})\label{fb}\\
\mbox{f}_{w}(q_{i-1},q_{i},d) & = & \mbox{score}_{\mathcal{M}_{w}}(d|q_{i-1},q_{i})\label{fw}
\end{eqnarray}

In total, the model uses three features and therefore needs a total
of three parameter weights referred to as $\lambda_{u}$, $\lambda_{b}$,
and $\lambda_{w}$. 

\subsection{Proof of the SDM Scoring Algorithm}

\begin{theorem} The SDM scoring algorithm as given in Equation \ref{eq:sdm-scoring}
implements the Markov random field as given in the factor graph of
Figure \ref{fig:Factor-graph}, with features defined as in Equations
\ref{fu}–\ref{fw}, and given parameters $\lambda_{u}$, $\lambda_{b}$,
and $\lambda_{w}$.

\end{theorem}

\begin{proof}

Starting with Equation \ref{eq:sdm-factorized} and using the choices
for factors and feature of Equations \ref{fu}–\ref{fw} yields

\[
\mbox{score}_{SDM}(d|\mathbf{q})\; \stackrel{rank}{=} \; \sum_{q_{i}\in\mathbf{q}}\mbox{f}_{u}(q_{i},d)\cdot\lambda_{u}+
\]


\[
\sum_{q_{i-1},q_{i}\in\mathbf{q}}\left(\mbox{f}_{b}(q_{i-1},q_{i},d)\cdot\lambda_{b}+\mbox{f}_{w}(q_{i-1},q_{i},d)\cdot\lambda_{w}\right)
\]

Reordering terms of the sums, and making use of the independence of
$\lambda$ from particular the query terms yields 

\[
\mbox{score}_{SDM}(d|\mathbf{q})\; \stackrel{rank}{=} \; \lambda_{u}\sum_{q_{i}\in\mathbf{q}}\mbox{f}_{u}(q_{i},d)+
\]

\[
\lambda_{b}\sum_{q_{i-1},q_{i}\in\mathbf{q}}\mbox{f}_{b}(q_{i-1},q_{i},d)+\lambda_{w}\sum_{q_{i-1},q_{i}\in\mathbf{q}}\mbox{f}_{w}(q_{i-1},q_{i},d)
\]


\begin{equation}
=\lambda_{u}\mbox{score}_{\mathcal{M}_{u}}(d|\mathbf{q})+\lambda_{b}\mbox{score}_{\mathcal{M}_{b}}(d|\mathbf{q})+\lambda_{w}\mbox{score}_{\mathcal{M}_{w}}(d|\mathbf{q})\label{eq:sdm-final}
\end{equation}

This is the SDM scoring equation given in Equation \ref{eq:sdm-scoring}.
\end{proof}

\subsection{Parameter Learning}

There are two common approaches to optimize settings of parameters
$\lambda$ for given relevance data: grid tuning or learning-to-rank.
Due to its low-dimensional parameter space, all combinations of choices
for $\lambda_{u}$, $\lambda_{b}$, and $\lambda_{w}$ in the interval
$(0,1)$ can be evaluated. For example a choice of 10 values leads
to 1000 combinations to evaluate. For rank equivalence, without loss
of generality it is sufficient to only consider nonnegative combinations where
$\lambda_{u}+\lambda_{b}+\lambda_{w}=1$, which reduces the number
of combinations to 100.

An alternative is to use a learning-to-rank algorithms such as using
coordinate ascent to directly optimize for a retrieval metric, e.g.
mean average precision (MAP). Coordinate ascent starts with an initial
setting, then continues to update one of the three dimensions in turn
to its best performing setting until convergence is reached. 

Since Equation \ref{eq:sdm-final} represents a log-linear model on
the three language models, any learning-to-rank algorithm including
Ranking SVM \cite{joachims2002ranksvm} can be used. However, in order
to prevent a mismatch between training phase and prediction phase
it is important to either use the whole collection to collect negative
training examples or to use the the same candidate selection strategy
(e.g., top 1000 documents under the unigram model) in both phases.
In this work, we use the RankLib\footnote{http://lemurproject.org/ranklib.php}
package in addition to grid tuning.

\section{Generative SDM Model}

\label{sec:genSDM}In this section we derive a generative model which
makes use of the same underlying unigram, bigram and windowed bigram
language models. Generative models are also called directed graphical
models or Bayesian networks. Generative models are often found to
be unintuitive, because the model describes a process that generates
data given variables we want to infer. In order to perform learning,
the inference algorithm 'inverts' the conditional relationships of
the process and to reason which input would most likely lead to the
observed data. 

\subsection{Generative Process: genSDM\label{subsec:Generative-process}}

We devise a generative model where the query and the document are
generated from distributions over unigrams $\phi_{d}^{u}$, over bigrams
$\phi_{d}^{b}$ and windowed bigrams $\phi_{d}^{w}$. These three
distributions are weighted according to a multinomial parameter $(\lambda_{u},\lambda_{b},\lambda_{w})$
of nonnegative entries that is normalized to sum to one.

The generative process is visualized in directed factor graph notation
\cite{dietz2010notation} in Figure \ref{fig:Generative-model}. For
a given document $d$ with according distributions, the query $\mathbf{q}=q_{1}q_{2}\dots q_{k}$
is assumed to be generated with the following steps:

\begin{itemize}[leftmargin=3mm]
\item Draw a multinomial distribution $\lambda$ over the set '$u$','$b$','$w$'.
\item Assume distributions to represent the document $d$ are given to model
unigrams $\phi_{d}^{u}$, bigrams $\phi_{d}^{b}$ and windowed bigrams
$\phi_{d}^{w}$.
\item Draw an indicator variable $Z\sim Mult(\boldsymbol{\lambda})$ to
indicate which distribution should be used.
\item If $Z=\mbox{'}u\mbox{'}$ then 

\begin{itemize}[leftmargin=3mm]
\item For all positions $1\leq i\leq k$ of observed query terms $q_{i}$
do: \\
Draw unigram $Q_{i}\sim Mult(\phi_{d}^{u})$.
\end{itemize}
\item If $Z=\mbox{'}b\mbox{'}$ then 

\begin{itemize}[leftmargin=3mm]
\item For all positions $2\leq i\leq k$ of observed query bigrams $q_{i-1},q_{i}$
do: \\
Draw bigram $(Q_{i-1},Q_{i})\sim Mult(\phi_{d}^{b})$.
\end{itemize}
\item If $Z=\mbox{'}w\mbox{'}$ then 

\begin{itemize}[leftmargin=3mm]
\item For all positions $2\leq i\leq k$ of observed query terms $q_{i-1},q_{i}$
do:\\
 Draw cooccurrence $\{Q_{i-1},Q_{i}\}\sim Mult(\phi_{d}^{w})$.
\end{itemize}
\end{itemize}
When scoring documents, we assume that parameters $\lambda_{u}$,
$\lambda_{b}$, and $\lambda_{w}$ are given and that the random variables
$Q_{i}$ are bound to the given query terms $q_{i}$. Furthermore,
the document representations $\phi_{d}^{u}$, $\phi_{d}^{b}$, $\phi_{d}^{w}$
are assumed to be fixed – we detail how they are estimated below. 

The only remaining random variables that remains is the draw of the
indicator $Z$. The probability of $Z$ given all other variables
being estimated in close form. E.g., $p(Z=\mbox{'}u\mbox{'}|\mathbf{q},\mbox{\ensuremath{\lambda}}\dots)\propto\lambda_{u}\prod_{i=1}^{k}\phi_{d}^{u}(q_{i})$
and analogously for 'b' and 'w', with a normalizer that equals the
sum over all three values.

Marginalizing (i.e., summing) over the uncertainty in assignments
of $Z$, this results as the following likelihood for all query terms
$\mathbf{q}$ under the generative model.

\begin{equation}
\mathcal{L}(\mathbf{q}|\lambda,\phi_{d}^{u},\phi_{d}^{b},\phi_{d}^{w})=\lambda_{u}\prod_{i=1}^{k}\phi_{d}^{u}(q_{i})+\lambda_{b}\prod_{i=2}^{k}\phi_{d}^{b}(\left(q_{i-1},q_{i}\right))+\lambda_{w}\prod_{i=2}^{k}\phi_{d}^{w}(\left\{ q_{i-1},q_{i}\right\} ) \label{eq:genmodel-likelihood}
\end{equation}

\subsection{Document Representation\label{subsec:Document-Representation-feature}}

In order for the generative process to be complete, we need to define
the generation for unigram, bigram and windowed bigram representations
of a document $d$. There are two common paradigms for bigram models,
the first is going back to n-gram models by generating word $w_{i}$
conditioned on the previous word $w_{i-1}$, where the other paradigm
is to perceive a document as a bag-of-bigrams which are drawn independently.
As the features of the sequential dependence model implement the latter
option, we focus on the bag-of-bigram approach here, and discuss the
n-gram approach in Section \ref{sec:genNgram}.

Each document $d$ in the corpus with words $w_{1},w_{2},\dots w_{n}$
is represented through three different forms. Each representation
is being used to model one of the multinomial distributions $\phi_{d}^{u}$,
$\phi_{d}^{b}$, $\phi_{d}^{w}$.

\paragraph{Bag of unigrams}

The unigram representation of $d$ follows the intuition of the document
as a bag-of-words $w_{i}$ which are generated independently through
draws from a multinomial distribution with parameter $\phi_{d}^{u}$. 

In the model, we further let the distribution $\phi_{d}^{u}$ be governed
by a Dirichlet prior distribution. In correspondence to the SDM model,
we choose the Dirichlet parameter that is proportional to the empirical
distribution in the corpus, i.e., $p(w)=\frac{n_{w,\star}}{n_{\star,\star}}$
with the scale parameter $\mu$. We denote this Dirichlet parameter
as $\tilde{\mu}^{u}=\left\{ \mu\cdot\frac{n_{w,\star}}{n_{\star,\star}}\right\} _{w\in V}$
which is a vector with entries for all words $w$ in the vocabulary
$V$.

The generative process for the unigram representation is:
\begin{enumerate}
\item Draw categorical parameter $\phi_{d}^{u}\sim Dir(\tilde{\mu}^{u})$.
\item For each word $w_{i}\in d$ do: \\
Draw $w_{i}\sim Mult(\phi_{d}^{u})$.
\end{enumerate}
Given a sequence of words in the document $d=w_{1}w_{2}\dots w_{n}$,
the parameter vector $\phi_{d}^{u}$ is estimated in closed form as
follows.

\[
\phi_{d}^{u}=\left\{ \frac{n_{w,d}+\mu\frac{n_{w,\star}}{n_{\star,\star}}}{n_{\star,d}+\mu}\right\} _{w\in V}
\]

The log likelihood of a given set of query terms $\mathbf{q}=q_{1}q_{2}\dots q_{k}$
under this model is given by

\[
\log\mathcal{L}_{u}(\mathbf{q}|\phi_{d}^{u})=\sum_{q_{i}\in\mathbf{q}}\log\frac{n_{q_{i},d}+\mu\frac{n_{q_{i},\star}}{n_{\star,\star}}}{n_{\star,d}+\mu}
\]

Notice, that $\log\mathcal{L}_{u}(\mathbf{q}|\phi_{d}^{u})$ is identical
to $\mbox{score}_{\mathcal{M}_{u}}(d|\mathbf{q})$ of Equation \ref{eq:unigram-model}. 

\paragraph{Bag of ordered bigrams}

One way of incorporating bigram dependencies in a model is through
a bag-of-bigrams representation. For a document $d$ with words $w_{1},w_{2},\dots w_{n}$
for every $i,\,2\leq i\leq n$ a bigram $(w_{i-1},w_{i})$ is placed
in the bag. The derivation follows analogously to the unigram case.
The multinomial distribution $\phi_{d}^{b}$ is drawn from a Dirichlet
prior distribution, parameterized by parameter $\tilde{\mu}^{b}$.
The Dirichlet parameter is derived from bigram-statistics from the
corpus, scaled by the smoothing parameter $\mu$.

The generative process for bigrams is as follows:
\begin{enumerate}
\item Draw categorical parameter $\phi_{d}^{b}\sim Dir(\tilde{\mu}^{b})$
\item For each pair of consecutive words $(w_{i-1},w_{i})\in d$: draw $(w_{i},w_{i+1})\sim Mult(\phi_{d}^{b})$
\end{enumerate}
Given an observed sequence of bigrams in the document $d=(w_{1},w_{2})(w_{2},w_{3})\dots(w_{n-1}w_{n})$,
the parameter vector $\phi_{d}^{b}$ can be estimated in closed form
as follows.

\[
\phi_{d}^{b}=\left\{ \frac{n_{(w,u),d}+\mu\frac{n_{(w,u),\star}}{n_{(\star,\star),\star}}}{n_{(\star,\star),d}+\mu}\right\} _{(w,u)\in VxV}
\]

The log likelihood of a given set of query terms $\mathbf{q}$ with\\
 $\mathbf{q}=(q_{1}q_{2}),(q_{2}q_{3})\dots(q_{k-1}q_{k})$
under this model is given by

\[
\log\mathcal{L}_{b}(\mathbf{q}|\phi_{d}^{b})=\sum_{(q_{i-1},q_{i})\in\mathbf{q}}\log\frac{n_{(q_{i-1},q_{i}),d}+\mu\frac{n_{(q_{i-1},q_{i}),\star}}{n_{(\star,\star),\star}}}{n_{(\star,\star),d}+\mu}
\]

Also, $\log\mathcal{L}_{b}(\mathbf{q}|\phi_{d}^{b})$ produces the
identical to $\mbox{score}_{\mathcal{M}_{b}}(d|\mathbf{q})$ above.

\paragraph{Bag of unordered windowed bigrams}

The windowed-bigram model of document $d$ works with a representation
of eight consecutive words $(w_{i-7}\dots w_{i})$, with derivation
analogously to the bigram case. However, in order to determine the
probability for two words $u$ and $v$ to occur within an unordered
window of 8 terms, we integrate over all positions and both directions.
The estimation of the windowed bigram parameter follows as

\[
\phi_{d}^{w}=\left\{ \frac{n_{\{u,v\}_{8},d}+\mu\frac{n_{\{u,v\}_{8},\star}}{n_{\{\star,\star\}_{8},\star}}}{n_{\{\star,\star\}_{8},d}+\mu}\right\} _{u\in V,v\in V}
\]

where $n_{\{u,v\}_{8},d}$ refers to the number of cooccurrences of
terms $u$ and $v$ within a window of eight terms. With parameters
$\phi_{d,v}^{w}$ estimated this way, the log-likelihood for query
terms $\mathbf{q}$ is given as

\begin{eqnarray*}
\log\mathcal{L}_{w}(\mathbf{q}|\phi_{d,\star}^{w}) & = & \sum_{{q_{i}\in\mathbf{q}\atop i>1}}\log\frac{n_{\left\{ q_{i-1},q_{i}\right\} _{8},d}+\mu\frac{n_{\left\{ q_{i-1},q_{i}\right\} _{8},\star}}{n_{\left\{ \star,\star\right\} _{8},\star}}}{n_{\left\{ q_{i-1},\star\right\} _{8},d}+\mu}
\end{eqnarray*}

The windowed bigram model $\mathcal{M}_{w}$ introduced above produces
the same score denoted $\mbox{score}_{\mathcal{M}_{w}}(d|\mathbf{q})$
as $\log\mathcal{L}_{w}(\mathbf{q}|\phi_{d}^{b})$.

\subsection{Generative Scoring Algorithm}

Inserting the expressions of the unigram, bigram and windowed bigram
language model into the likelihood of the generative model (Equation
\ref{eq:genmodel-likelihood}), yields 

\begin{eqnarray}
\mathcal{L}_{\mbox{Gen}}(\mathbf{q},d) & \propto & \lambda_{u}\exp\,\mbox{score}_{\mathcal{M}_{u}}(d|\mathbf{q})\label{eq:likelihood-genSDM}
\end{eqnarray}

\[
+\lambda_{b}\exp\,\mbox{score}_{\mathcal{M}_{b}}(d|\mathbf{q})+\lambda_{w}\exp\,\mbox{score}_{\mathcal{M}_{w}}(d|\mathbf{q})
\]

Since the expressions such as $\prod_{i=1}^{k}\phi_{d}^{u}(q_{i})$
are identical to \\
 $\exp\,\mbox{score}_{\mathcal{M}_{u}}(d|\mathbf{q})$
as it was introduced in Section \ref{sec:Sequential-Dependence-Algo}.

\subsection{Connection to MRF-SDM model}

We want to point out the similarity of the likelihood of the generative
SDM model (Equation \ref{eq:likelihood-genSDM}) and the log-likelihood
of the SDM Markov random field from Equation \ref{eq:sdm-final},
which (as a reminder) is proportional to 

\begin{eqnarray}
\log\mathcal{L}_{\mbox{MRF}}(\mathbf{q},d) & \propto & \lambda_{u}\mbox{score}_{\mathcal{M}_{u}}(d|\mathbf{q})+\lambda_{b}\mbox{score}_{\mathcal{M}_{b}}(d|\mathbf{q})+\lambda_{w}\mbox{score}_{\mathcal{M}_{w}}(d|\mathbf{q})\label{eq:recap-log-mrf-sdm}
\end{eqnarray}

The difference between both likelihood expressions is that for MRF,
the criterion is optimized in log-space (i.e., $\underline{\log}\mathcal{L}_{\mbox{MRF}}(\mathbf{q},d)$)
where for the generative model, the criterion is optimized in the
space of probabilities (i.e., $\mathcal{L}_{\mbox{Gen}}(\mathbf{q},d)$).
Therefore the MRF is optimizing a linear-combination of log-features
such as $\mbox{score}_{\mathcal{M}_{u}}(d|\mathbf{q})$, where by
contrast, the generative model optimizes a linear combination of probabilities
such as $\underline{\exp}\,\mbox{score}_{\mathcal{M}_{u}}(d|\mathbf{q})$. 

Looking at Equation \ref{eq:recap-log-mrf-sdm} in the probability
space, it becomes clear that the weight parameter $\lambda$ acts
on the language models through the exponent (and not as a mixing factor):

\begin{eqnarray*}
\mathcal{L}_{\mbox{MRF}}(\mathbf{q},d) & \propto & \left(\exp\,\mbox{score}_{\mathcal{M}_{u}}(d|\mathbf{q})\right)^{\lambda_{u}} \cdot\left(\exp\,\mbox{score}_{\mathcal{M}_{b}}(d|\mathbf{q})\right)^{\lambda_{b}}\cdot\left(\exp\,\mbox{score}_{\mathcal{M}_{w}}(d|\mathbf{q})\right)^{\lambda_{w}}
\end{eqnarray*}

This difference is the reason why the MRF factor functions are called log-linear
models and why the parameter $\boldsymbol{\lambda}$ is not restricted
to nonnegative entries that sum to one---although this restriction
can be imposed to restrict the parameter search space without loss
of generality. 

\subsection{Connections to Jelinek-Mercer Smoothing}

Jelinek-Mercer smoothing \cite{chen1996smoothing} is an interpolated
language smoothing technique. While discussed as an alternative to
Dirichlet smoothing by Zhai et al.\ \cite{zhai2001smoothing}, here
we analyze it as a paradigm to combine unigram, bigram, and windowed
bigram model.

The idea of Jelinek-Mercer smoothing is to combine a complex model
which may suffer from data-sparsity issues, such as the bigram language
model, with a simpler back-off model. Both models are combined by
linear interpolation.

We apply Jelinek-Mercer smoothing to our setting through a nested
approach. The bigram model is first smoothed with a windowed bigram
model as a back-off distribution with interpolation parameter $\tilde{\lambda}_{b}$.
Then the resulting model is smoothed additionally with a unigram model
with parameter $\tilde{\lambda}_{u}$. This model results in the following
likelihood for optimization.

\[
\mathcal{L}_{\mbox{JM}}(\mathbf{q},d)\propto(1-\tilde{\lambda}_{u})\Bigl(\tilde{\lambda}_{b}\exp\,\mbox{score}_{\mathcal{M}_{b}}(d|\mathbf{q})+ (1-\tilde{\lambda}_{b})\exp\,\mbox{score}_{\mathcal{M}_{w}}(d|\mathbf{q})\Bigr)+(\tilde{\lambda}_{u})\exp\,\mbox{score}_{\mathcal{M}_{u}}(d|\mathbf{q})
\]

We demonstrate that this function is equivalent to the likelihood
of the generative model (Equation \ref{eq:likelihood-genSDM}), through
the reparametrization of $\lambda_{u}=\tilde{\lambda}_{u}$, $\lambda_{b}=(1-\tilde{\lambda}_{u})\cdot\tilde{\lambda}_{b}$
and $\lambda_{w}=(1-\tilde{\lambda}_{u})\cdot(1-\tilde{\lambda}_{b})$.
Therefore, we conclude that the generative model introduced in this
section is equivalent to a Jelinek-Mercer-smoothed bigram model discussed
here. 

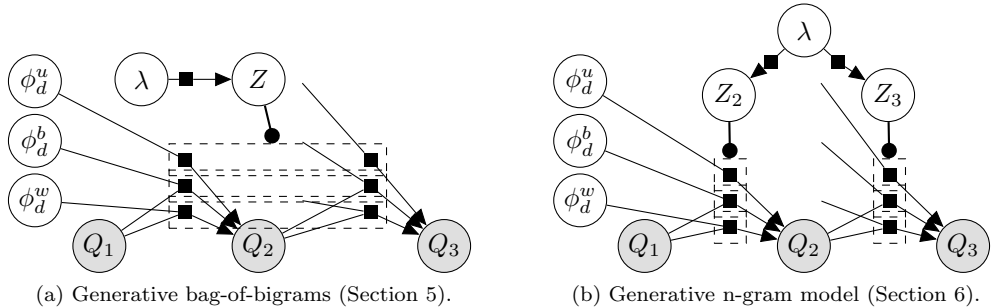
\begin{figure*}
\hfill{} \subfloat[Generative bag-of-bigrams (Section \ref{sec:genSDM}).\label{fig:Generative-bof-model}]{\begin{tikzpicture}[x=0.7cm,y=0.5cm]


\node[latent] (phiu)	{$\phi^u_d$} ; %
\node[latent, below=0.15 of phiu] (phib)	{$\phi^b_d$} ; %
\node[latent, below=0.15 of phib] (phiw)	{$\phi^w_d$} ; %

\node[obs, below right=0.2 and 0.5 of phiw]    (Q1)      {$Q_1$} ; %
\node[obs, right=2 of Q1]    (Q2)      {$Q_2$} ; %
\node[obs, right=2.5 of Q2]    (Q3)      {$Q_3$} ; %

\node[latent, above=3 of Q2]    (Z2)      {$Z$} ; %

\node[latent, left=1.2 of Z2]	(lambda)	{$\lambda$} ; %

\node[const, above left=-0.3 and 1 of Q2] (marker) {} ; %

\factor[above=1.8 of marker]     {q2u}     {} {} {} ; %
\factor[above=1.1 of marker]     {q2b}     {} {} {} ; %
\factor[above=0.4 of marker]     {q2w}     {} {} {} ; %

\factoredge {phiu}  {q2u}  {Q2} ; %
\factoredge {phib}  {q2b}  {Q2} ; %
\factoredge {phiw}  {q2w}  {Q2} ; %

\draw (Q1) -- (q2b) ; %
\draw (Q1) -- (q2w) ; %

\node[const, above left=-0.3 and 1 of Q3] (marker2) {} ; %

\factor[above=1.8 of marker2]     {q3u}     {} {} {} ; %
\factor[above=1.1 of marker2]     {q3b}     {} {} {} ; %
\factor[above=0.4 of marker2]     {q3w}     {} {} {} ; %
\factor[right=0.2 of lambda]	{lambda2}	{} {} {} ;%

\node[const, right=4.5 of phiu] (phiu2) {}; %
\node[const, right=4.5 of phib] (phib2) {}; %
\node[const, right=4.5 of phiw] (phiw2) {}; %

\factoredge {phiu2}  {q3u}  {Q3} ; %
\factoredge {phib2}  {q3b}  {Q3} ; %
\factoredge {phiw2}  {q3w}  {Q3} ; %

\draw (Q2) -- (q3b) ; %
\draw (Q2) -- (q3w) ; %

\factoredge {lambda}	{lambda2} 	{Z2} ; %

\gate  {z2gate1} {(q2u) (q3u)} {Z2} ; %
\gate  {z2gate2} {(q2b) (q3b)} {} ; %
\gate  {z2gate3} {(q2w) (q3w)} {} ; %

\end{tikzpicture}

} \hfill{}\subfloat[Generative n-gram model (Section \ref{sec:genNgram}).\label{fig:Generative-ngram-model}]{\begin{tikzpicture}[x=0.7cm,y=0.5cm]


\node[latent] (phiu)	{$\phi^u_d$} ; %
\node[latent, below=0.15 of phiu] (phib)	{$\phi^b_d$} ; %
\node[latent, below=0.15 of phib] (phiw)	{$\phi^w_d$} ; %

\node[obs, below right=0.2 and 0.5 of phiw]    (Q1)      {$Q_1$} ; %
\node[obs, right=2 of Q1]    (Q2)      {$Q_2$} ; %
\node[obs, right=2 of Q2]    (Q3)      {$Q_3$} ; %

\node[latent, above left=3 and 0.7 of Q2]    (Z2)      {$Z_2$} ; %
\node[latent, above left=3 and 0.7 of Q3]    (Z3)      {$Z_3$} ; %

\node[latent, above right=1 of Z2]	(lambda)	{$\lambda$} ; %

\node[const, above left=-0.3 and 1 of Q2] (marker) {} ; %

\factor[above=1.4 of marker]     {q2u}     {} {} {} ; %
\factor[above=0.7 of marker]     {q2b}     {} {} {} ; %
\factor[above=0 of marker]     {q2w}     {} {} {} ; %

\factoredge {phiu}  {q2u}  {Q2} ; %
\factoredge {phib}  {q2b}  {Q2} ; %
\factoredge {phiw}  {q2w}  {Q2} ; %

\draw (Q1) -- (q2b) ; %
\draw (Q1) -- (q2w) ; %

\node[const, above left=-0.3 and 1 of Q3] (marker2) {} ; %

\factor[above=1.4 of marker2]     {q3u}     {} {} {} ; %
\factor[above=0.7 of marker2]     {q3b}     {} {} {} ; %
\factor[above=0 of marker2]     {q3w}     {} {} {} ; %
\factor[below left=0.2 of lambda]	{lambda2}	{} {} {} ;%
\factor[below right=0.2 of lambda]	{lambda3}	{} {} {} ;%

\node[const, right=4 of phiu] (phiu2) {}; %
\node[const, right=4 of phib] (phib2) {}; %
\node[const, right=4 of phiw] (phiw2) {}; %

\factoredge {phiu2}  {q3u}  {Q3} ; %
\factoredge {phib2}  {q3b}  {Q3} ; %
\factoredge {phiw2}  {q3w}  {Q3} ; %

\draw (Q2) -- (q3b) ; %
\draw (Q2) -- (q3w) ; %

\factoredge {lambda}	{lambda2} 	{Z2} ; %
\factoredge {lambda}	{lambda3} 	{Z3} ; %

\gate  {z2gate1} {(q2u)} {Z2} ; %
\gate  {z2gate2} {(q2b)} {} ; %
\gate  {z2gate3} {(q2w)} {} ; %

\gate  {z3gate1} {(q3u)} {Z3} ; %
\gate  {z3gate2} {(q3b)} {} ; %
\gate  {z3gate3} {(q3w)} {} ; %

\end{tikzpicture}

}\hfill{} 

\caption{Generative n-gram mixture models.}
\end{figure*}

\section{Generative N-Gram-based Model}

\label{sec:genNgram}The generative model introduced in Section \ref{sec:genSDM}
is rather untypical in that it considers three bag-of-features representations
of a single document without ensuring consistency among them. Using
it to generate documents might yield representations of different
content. In this section we discuss a more stereotypical generative
model based on the n-gram process (as opposed to a bag-of-n-grams).
Consistently with previous sections, this model combines a unigram,
bigram, and windowed bigram model.

While the unigram model is exactly as described in Section \ref{subsec:Document-Representation-feature},
the setup for the bigram and windowed bigram cases change significantly
when moving from a bag-of-bigram paradigm to an n-gram paradigm. 

\subsection{Generative N-gram-based Bigram Process\label{subsec:Generative-n-gram-style-bigram}}

In the bag-of-bigrams model discussed in Section \ref{subsec:Document-Representation-feature},
both words of a bigram $(w_{i-1},w_{i})$ are drawn together from
one distribution $\phi_{d}$ per document $d$. In contrast, in the
n-gram models we discuss here, $w_{i}$ is drawn from a distribution
that is conditioned on $w_{i-1}$ in addition to $d$, i.e., $\phi_{d,w_{i-1}}$.
The difference is that where in the bag-of-bigrams model follows $p(w,v|d)=\frac{n_{(v,w),d}}{n_{(\star,\star),d}}$,
the n-gram version follows $p(w|v,d)=\frac{n_{(v,w),d}}{n_{(v,\star),d}}$
.

As before, we use language models with Dirichlet smoothing, a smoothing
technique that integrates into the theoretical generative framework
through prior distributions. For all terms $v\in V$, we let each
language model $\phi_{d,v}$ be drawn from a Dirichlet prior with
parameter $\tilde{\mu}_{v}^{b}$, which is based on bigram statistics
from the corpus, which are scaled by the smoothing parameter $\mu$.
For bigram statistics, we have the same choice between a bag-of-bigram
and n-gram paradigm. For consistency we choose to follow the n-gram
paradigm which yields Dirichlet parameter $\tilde{\mu}_{v}^{b}=\left\{ \mu\frac{n_{(v,w),\star}}{n_{(v,\star),\star}}\right\} _{w\in V}$
.

The generative process for the bigram model is as follows:
\begin{enumerate}
\item For all words $v\in V$ in the vocabulary: draw categorical parameter
$\phi_{d,v}^{b}\sim Dir(\tilde{\mu}_{v}^{b})$.
\item Draw the first word of the document $w_{1}\in d$ from the unigram
distribution, $w_{1}\sim Mult(\phi_{d}^{u})$.
\item For each remaining word $w_{i}\in d;\:i\leq2$: \\
draw $w_{i}\sim Mult(\phi_{d,w_{i-1}}^{b})$.
\end{enumerate}
Given a sequence of words in the document $d=w_{1}w_{2}\dots w_{n}$,
the parameter vectors $\phi_{d,v}^{b}$ ($\forall v\in V$) can be
estimated in closed form as follows.
\[
\phi_{d,v}^{b}=\left\{ \frac{n_{(v,w),d}+\mu\frac{n_{(v,w),\star}}{n_{(v,\star),\star}}}{n_{(v,\star),d}+\mu}\right\} _{w\in V}
\]

The log likelihood of a given set of query terms $\mathbf{q}=q_{1}q_{2}\dots q_{k}$
is modeled as $p(\mathbf{q})=\left(\prod_{{q_{i}\in\mathbf{q}\atop i>1}}p(q_{i}|q_{i-1})\right)\cdot p(q_{1})$.
With parameters $\phi_{d,v}^{b}$ as estimated above, the log-likelihood
for query terms $\mathbf{q}$ is given as

\begin{eqnarray*}
\log\mathcal{L}_{b}(\mathbf{q}|\phi_{d,\star}^{b}) & = & \sum_{{q_{i}\in\mathbf{q}\atop i>1}}\log\frac{n_{(q_{i-1},q_{i}),d}+\mu\frac{n_{(q_{i-1},q_{i}),\star}}{n_{(q_{i-1},\star),\star}}}{n_{(q_{i-1},\star),d}+\mu}  +\log\mathcal{L}_{u}(q_{1}|\phi_{d}^{u})
\end{eqnarray*}

The second term handles the special case of the first query word $q_{1}$ which has
no preceding terms and therefore, when marginalizing over all possible preceding terms,  collapses to the unigram distribution.

Even when ignoring the special treatment for the first query term
$q_{1}$, the bigram model $\mathcal{M}_{b}$ referred to above as
$\mbox{score}_{\mathcal{M}_{b}}(d|\mathbf{q})$ produces the different
score as $\log\mathcal{L}_{b}(\mathbf{q}|\phi_{d}^{b})$ due to the
difference in conditional probability and joint probability.

\subsection{Generative Windowed-Bigram Process\label{subsec:Generative-windowed-bigram-proce}}

The windowed bigram model of document $d$ also represents each word
$w_{i}$ as a categorical distribution. The difference is that the
model conditions on a random word within the 8-word window surrounding
 the $i$'th position. This is modeled by a random
draw of a position $j$ to select the word $w_{j}$ on which the draw
of word $w_{i}$ will be conditioned on. In the following, we denote
the set of all words surrounding word $w_{i}$ by $\omega_{i}=\{w_{i-7}\dots w_{i-1}w_{i+1}\dots w_{i+7}\}$.

The generative process for the windowed bigram model is as follows:
\begin{enumerate}[leftmargin=4mm]
\item For all words $v\in V$: draw categorical parameter $\phi_{d,v}^{w}\sim Dir(\tilde{\mu}_{v}^{w})$.
\item For each word $w_{i}\in d$: 

\begin{enumerate}[leftmargin=4mm]
\item Draw an index $j$ representing word $w_{j}\in\omega_{i}$ uniformly
at random.
\item Draw $w_{i}\sim Mult(\phi_{d,w_{j}}^{w})$.
\end{enumerate}
\end{enumerate}
Deriving an observed sequence of windows $\omega_{1}\omega_{2}\dots\omega_{n}$
from an given sequence of words in the document $d=w_{1}w_{2}\dots w_{n}$.
The parameter vectors $\phi_{d,v}^{w}$ ($\forall v\in V$) can be
estimated in closed form by counting all co-occurrences of $w_{i}$
with $v\in\omega_{i}$ in the vocabulary $V$. This quantity was introduced
above as $n_{\left\{ w,v\right\} _{8},d}$. In order to incorporate
choosing the position $j$, the co-occurrence counts are weighted
by the domain size of the uniform draw, i.e., $\frac{1}{7+7}$. 

\begin{eqnarray*}
\phi_{d,v}^{w} & = & \left\{ \frac{\frac{1}{14}n_{\left\{ v,w\right\} _{8},d}+\mu\frac{\frac{1}{14}n_{\left\{ v,w\right\} _{8},\star}}{\frac{1}{14}n_{\left\{ v,\star\right\} _{8},\star}}}{\frac{1}{14}n_{\left\{ v,\star\right\} _{8},d}+\mu}\right\} _{w\in V}
\end{eqnarray*}

As the factors $\frac{1}{14}$ cancel, we arrive at the second line.

With parameters $\phi_{d,v}^{w}$ as estimated above, the log-likelihood
for query terms $\mathbf{q}$ is given as

\[
\log\mathcal{L}_{w}(\mathbf{q}|\phi_{d,\star}^{w})=\sum_{{q_{i}\in\mathbf{q}\atop i>1}}\log\frac{n_{\left\{ q_{i-1},q_{i}\right\} _{8},d}+{\scriptstyle 14\cdot\mu}\cdot\frac{n_{\left\{ q_{i-1},q_{i}\right\} _{8},\star}}{n_{\left\{ q_{i-1},\star\right\} _{8},\star}}}{n_{\left\{ q_{i-1},\star\right\} _{8},d}+14\mu} +\log\mathcal{L}_{u}(q_{1}|\phi_{d}^{u})
\]

The second term handles the special case of the $q_{1}$ which has
no preceding terms and collapses to the unigram model.

Aside from the special treatment for $q_{1}$, the bigram model $\mathcal{M}_{w}$
introduced above $\mbox{score}_{\mathcal{M}_{w}}(d|\mathbf{q})$ produces
a different log score as $\log\mathcal{L}_{w}(\mathbf{q}|\phi_{d}^{b})$.

\subsection{A New Generative Process: genNGram}

The n-gram paradigm language models discussed in this section, allows
to generate a term $q_{i}$ optionally conditioned on the previous
term. This allows to integrate unigram, bigram, and windowed bigram
models with term-dependent choices. For instance, after generating
$q_{1}$ from the unigram model, $q_{2}$ might be generated from
a bigram model (conditioned on $q_{1}$), and $q_{3}$ generated from
the windowed bigram model (conditioned on $q_{2}$). These term-by-term
model choices are reflected in a list of latent indicator variables
$Z_{i}$, one for each query term position $q_{i}$.

The generative process is as follows.
\begin{itemize}[leftmargin=3mm]
\item Draw a multinomial distribution $\boldsymbol{\lambda}$ over the set
'$u$','$b$','$w$'.
\item Assume estimated unigram model $\phi_{d}^{u}$, bigram model $\phi_{d,v}^{b};\forall v\in V$
and windowed bigram model $\phi_{d,v}^{w};\forall v\in V$ that represent
the document $d$ as introduced in this section.
\item For the first query term $q_{1}$ do: \\
Draw $Q_{1}\sim Mult(\phi_{d}^{u})$.
\item For all positions $2\leq i\leq k$ of query terms $q_{i}$, do:

\begin{itemize}[leftmargin=3mm]
\item Draw an indicator variable $Z_{i}\sim Mult(\boldsymbol{\lambda})$
to indicate which distribution should be used.
\item If $Z_{i}=\mbox{'}u\mbox{'}$ then do: \\
Draw $Q_{i}\sim Mult(\phi_{d}^{u})$
from the unigram model (Section \ref{subsec:Document-Representation-feature}).
\item If $Z_{i}=\mbox{'}b\mbox{'}$ then do: \\
Draw $Q_{i}\sim Mult(\phi_{d,Q_{i-1}}^{b})$
from the bigram model (Section \ref{subsec:Generative-n-gram-style-bigram}).
\item If $Z_{i}=\mbox{'}w\mbox{'}$ then do:\footnote{In spirit with SDM, $\phi^{w}$ is estimated from eight-term windows
in the document, but only the previous word is considered when generating
the query.} Draw $Q_{i}\sim Mult(\phi_{d,Q_{i-1}}^{w})$ from the windowed bigram
model (Section \ref{subsec:Generative-windowed-bigram-proce}).
\end{itemize}
\end{itemize}
Assuming that all variables $Q_{i}$ and parameters $\phi$, $\lambda$
are given, only the indicator variables $Z_{i}$ need to be estimated.
Since all $Z_{i}$ are conditionally independent when other variables
are given, their posterior distribution can be estimated in closed-form.
For instance, $p(Z_{i}=\mbox{'}b\mbox{'}|\mathbf{q},\mbox{\ensuremath{\lambda}}\dots)\propto\lambda_{b}\phi_{d,q_{i-1}}^{b}(q_{i})$
and analogously for 'u' and 'w'.

Integrating out the uncertainty in $Z_{i}$ and considering all query
terms $q_{i}$, the model likelihood is estimated as

\begin{eqnarray}
\mathcal{L}(\mathbf{q}|\lambda,\phi_{d}^{u},\phi_{d}^{b},\phi_{d}^{w}) & = & \phi_{d}^{u}(q_{1})\cdot\prod_{i=2}^{k}\biggl(\lambda_{u}\phi_{d}^{u}(q_{i}) +\lambda_{b}\phi_{d,q_{i-1}}^{b}(q_{i})+\lambda_{w}\phi_{d,q_{i-1}}^{w}(q_{i})\biggr)\label{eq:Likelihood-genNGram}
\end{eqnarray}

\section{Experimental Evaluation}

\label{sec:Experiments}In this section, the theoretical analysis
of the family of dependency models is complemented with an empirical
evaluation. The goal of this evaluation is to understand implications
of different model choices in isolation. 

\begin{figure*}
\noindent \begin{centering}
\subfloat[Different methods and features.\label{fig:methods}]{\noindent \begin{centering}
\begin{tikzpicture}[x=0.5cm,y=0.3cm,  every node/.style={scale=0.7}]


\node (bowb)	{bag of windowed bigrams}; %
\node[below=1 of bowb] (bob)	{bag of bigrams}; %
\node[below=1 of bob] (uni)	{unigram model}; %
\node[below=1 of uni] (big)	{n-gram bigram model}; %
\node[below=1 of big] (wbig)	{n-gram windowed bigram}; %


\node[left=2 of bob, draw] (mrfSDM) {mrfSDM}; %
\node[below=3 of mrfSDM, draw] (mrfNGram) {mrfNGram}; %
\node[right=2.5 of uni, draw] (QL) {QL}; %

\node[right=2 of bob, draw] (genSDM) {genSDM}; %
\node[below=3 of genSDM, draw] (genNGram) {genNGram}; %

\draw[->] (mrfSDM) -> (bob); %
\draw[->] (mrfSDM) -> (bowb); %
\draw[->] (mrfSDM) -> (uni); %

\draw[->] (genSDM) -> (bob); %
\draw[->] (genSDM) -> (bowb); %
\draw[->] (genSDM) -> (uni); %

\draw[->] (QL) -> (uni); %

\draw[->] (mrfNGram) -> (big); %
\draw[->] (mrfNGram) -> (wbig); %
\draw[->] (mrfNGram) -> (uni); %

\draw[->] (genNGram) -> (big); %
\draw[->] (genNGram) -> (wbig); %
\draw[->] (genNGram) -> (uni); %

\node[above=2 of mrfSDM] (MRF)  {\textbf{MRF}} ; %
\node[above=2 of genSDM] (GEN)  {\textbf{Generative}} ; %


\node[below=1 of wbig] (base)	{}; %

\end{tikzpicture}
\par\end{centering}
}\subfloat[Performance with grid tuning.\label{fig:results-grid}]{\includegraphics[width=0.3\columnwidth]{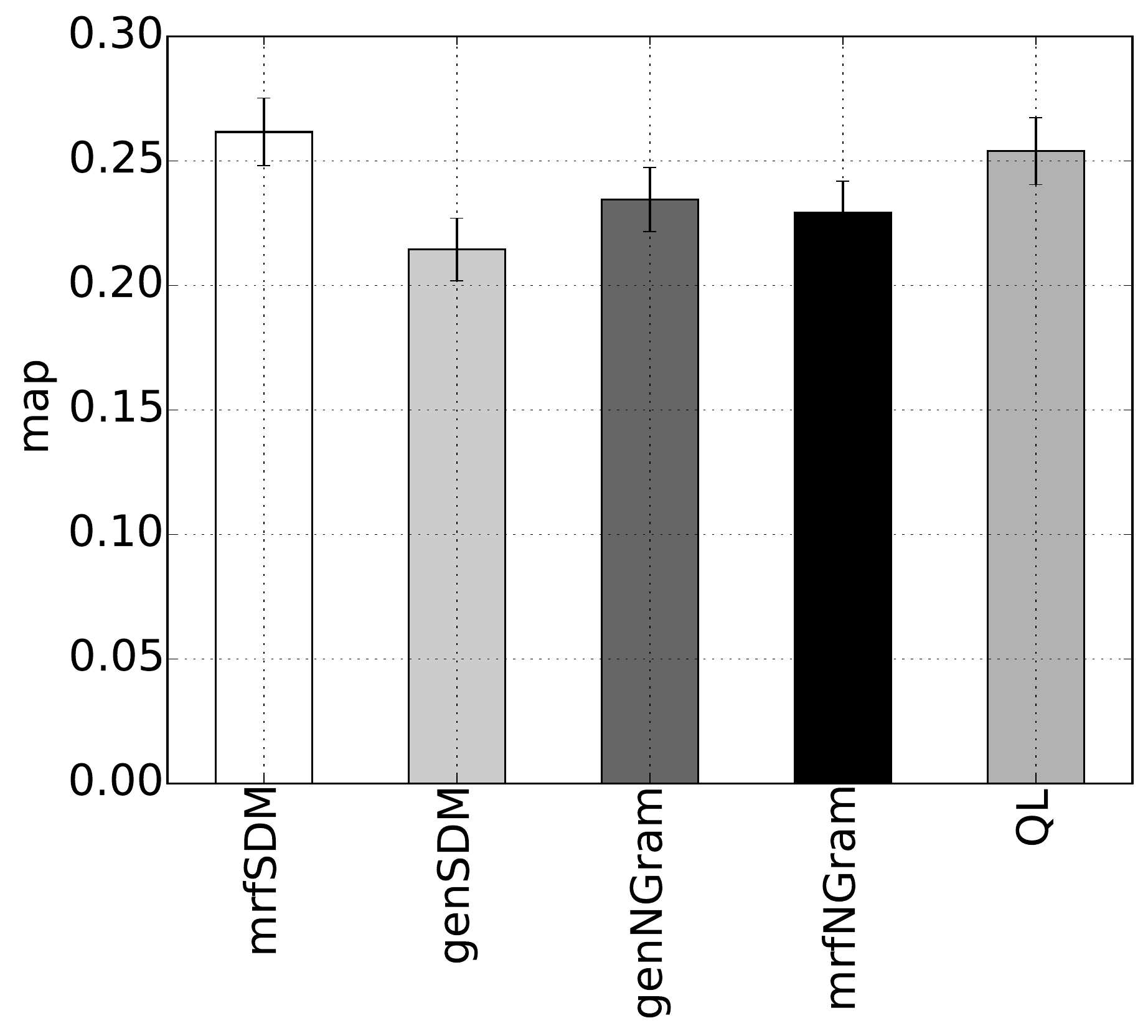}

}\subfloat[Performance with RankLib.\label{fig:results-ranklib}]{\includegraphics[width=0.3\columnwidth]{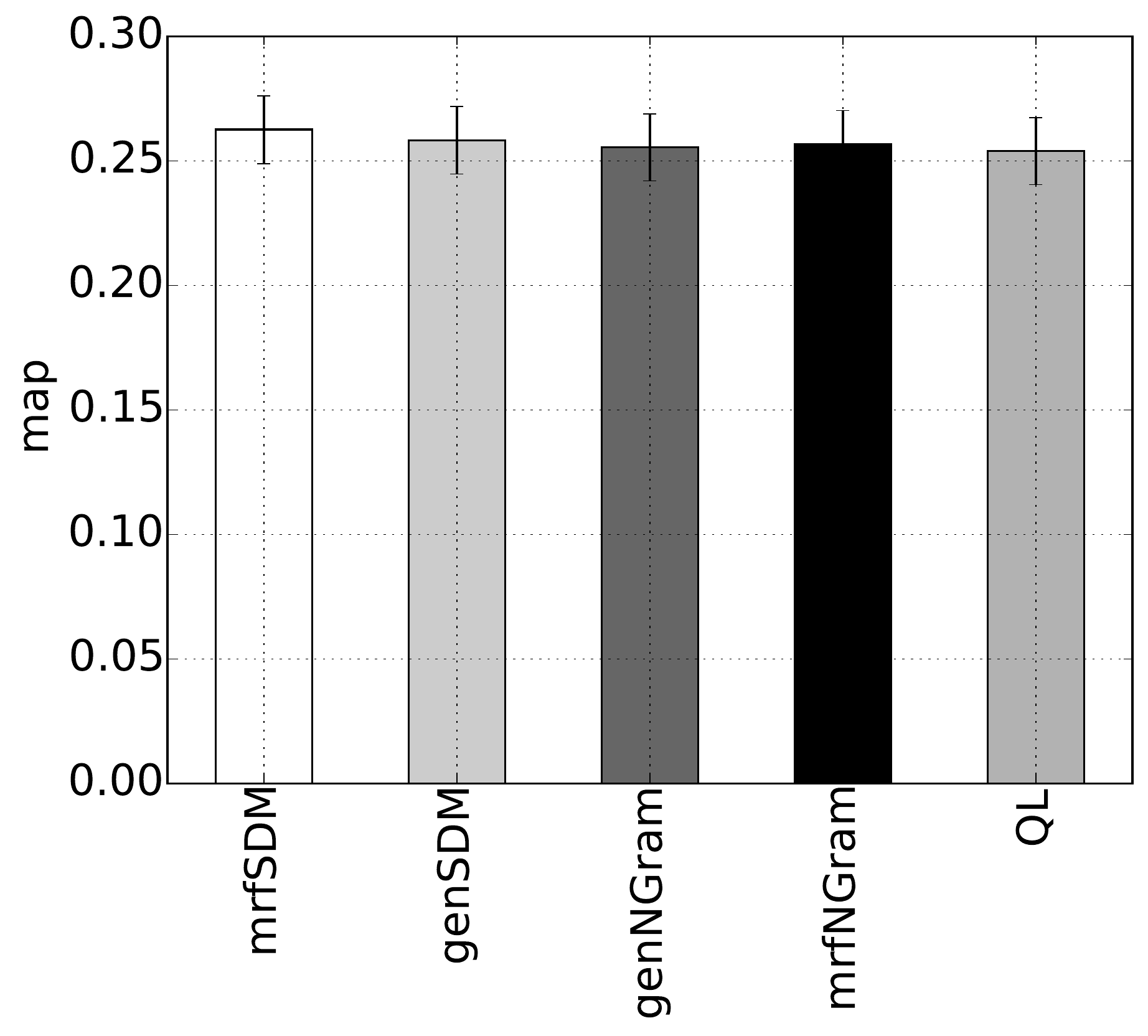}

}
\par\end{centering}
\caption{Experimental evaluation and results\label{fig:experiments}}
\end{figure*}
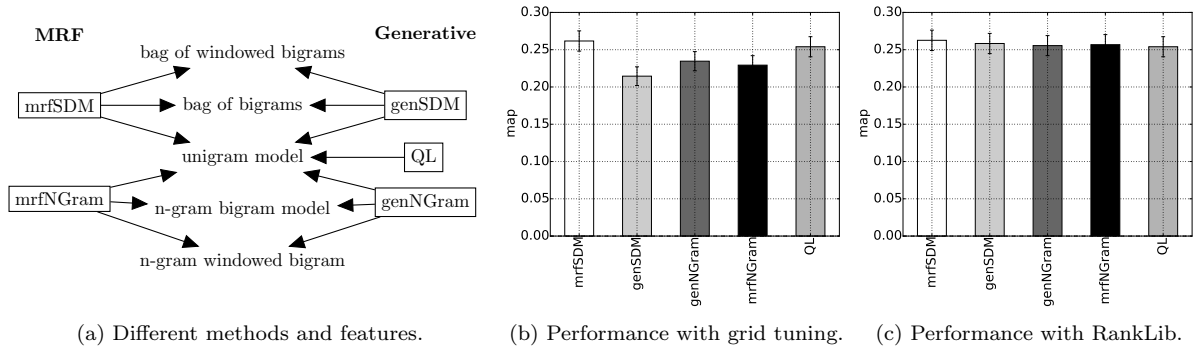

We compare the MRF-based and generative models with both paradigms
for bigram models. In particular, the following methods are compared
(cf. Figure \ref{fig:methods}):
\begin{itemize}[leftmargin=5mm]
\item \textbf{mrfSDM:} The original MRF-based sequential dependence model as introduced
by Metzler et al.\ \cite{metzler2005sdm}, as described in Section \ref{sec:mrfSDM}.
\item \textbf{genSDM:} A generative model with the same features, using the bag-of-bigrams
approach introduced in Section \ref{sec:genSDM}.
\item \textbf{genNGram:} Alternative generative model with using conditional bigram
models, closer to traditional n-gram models, discussed in Section
\ref{sec:genNgram}.
\item \textbf{mrfNGram:} A variant of the MRF-based SDM model using features from
conditional bigram models.
\item \textbf{QL:} The query likelihood model with Dirichlet smoothing, which is called the unigram
model in this paper.
\end{itemize}

All underlying language models are smoothed with Dirichlet smoothing,
as a preliminary study with Jelinek Mercer smoothing yielded worse
results. (This finding is consistent with a study of Smucker et al.~\cite{smucker2006smoothing}.)

Term probabilities of different language models are on very different
scales. Such as is the average probability of bag-of-bigram entry
is much smaller than a probability under the unigram model, which
is in turn much smaller than a term under a conditional bigram model.
As we anticipate that the Dirichlet scale parameter $\mu$ needs to be
adjusted we introduce separate parameters for different
language models (and not use parameter tying).

\subsection{Experimental Setup}

Aiming for a realistic collection with rather complete assessments
and multi-word queries, we study method performance on the Robust04
test set. The test set contains 249 queries\footnote{Removing query 672 which does not contain positive judgments. }
and perform tokenization on whitespace, stemming with Krovetz stemmer,
but only remove stopwords for unigram models. While we focus on the measure mean-average
precision (MAP), similar results are obtained for ERR@20, R-Precision,
bpref, MRR, and P@10 (available upon request).

We use five-fold cross validation using folds that are identical to
empirical studies of Huston et al.~\cite{huston2013termdependencies,huston2014termdepedencies-appendix}.
The training fold is used to select both the Dirichlet scale parameters
$\mu$ and weight parameters $\boldsymbol{\lambda}$. Performance
is measured on the test fold only.

Parameters are estimated in two phases. First the Dirichlet scale
parameter $\mu$ is selected to maximize retrieval performance (measured
in MAP) of each language model individually. See Table \ref{tab:tuning-mu}
for range of the search grid, estimated Dirichlet parameter,
and training performance.

In the subsequent phase, Dirichlet parameters are held fixed while
the weight parameter $\boldsymbol{\lambda}=\{\lambda_{u},\lambda_{b},\lambda_{w}\}$
is selected. To avoid performance differences due different machine
learning algorithms, we evaluate two learning approaches for weight
parameter $\boldsymbol{\lambda}$: grid search and coordinate ascent
from RankLib. Despite not strictly being necessary, for grid search we only consider nonnegative weights
that sum to one, as suggested in the original SDM paper \cite{metzler2005sdm}. Each weight entry is selected on
a grid $\lambda\in[0.0,0.05,\ldots0.95,1.0]$ while constraint-violating
combinations are discarded. The RankLib experiment does not use a
grid, but performs coordinate-ascent with five restarts.

For single-term queries, all discussed approaches reduce to the Query
Likelihood model, i.e., unigram model. We therefore hold them out
during the training phase, but include them in the test phase, where
they obtain the same ranking for all approaches.

\begin{table}
\caption{Dirichlet settings with max MAP on the train set.\label{tab:tuning-mu}}

\hfill{}\subfloat[Bag-of-bigram models.]{
\noindent \centering{}{}%
\begin{tabular}{c@{\hskip 3em}ll@{\hskip 3em}ll@{\hskip 3em}ll}
\toprule
{split} & {$\mu_{u}$} & {MAP} & {$\mu_{b}$} & {MAP} & {$\mu_{w}$} & {MAP}\tabularnewline
\midrule
{0} & {1000} & {0.252} & {18750} & {0.131} & {20000} & {0.171}\tabularnewline
{1} & {1000} & {0.253} & {18750} & {0.127} & {2500} & {0.163}\tabularnewline
{2} & {1000} & {0.252} & {18750} & {0.131} & {20000} & {0.165}\tabularnewline
{3} & {1000} & {0.254} & {18750} & {0.135} & {20000} & {0.168}\tabularnewline
{4} & {1000} & {0.259} & {21250} & {0.130} & {2500} & {0.170}\tabularnewline
\bottomrule
\multicolumn{7}{c}{{$\mu\in[10,250,500,\ldots,2500,3000,3500,\ldots,5000,10000]$}}
\end{tabular}{\scriptsize \par}}\hfill{}

\hfill{}\subfloat[N-gram models.\label{tab:genWindowedBigram}]{
\noindent \centering{}{}%
\begin{tabular}{c@{\hskip 3em}ll@{\hskip 3em}ll@{\hskip 3em}ll}
\toprule 
{split} & {$\mu_{u}$} & {MAP} & {$\mu_{b}$} & {MAP} & {$\mu_{w}$} & {MAP}\tabularnewline
\midrule
{0} & {1000} & {0.252} & {5} & {0.171} & {1} & {0.213}\tabularnewline
{1} & {1000} & {0.253} & {5} & {0.172} & {1} & {0.209}\tabularnewline
{2} & {1000} & {0.252} & {5} & {0.168} & {1} & {0.206}\tabularnewline
{3} & {1000} & {0.254} & {5} & {0.175} & {1} & {0.210}\tabularnewline
{4} & {1000} & {0.259} & {5} & {0.172} & {1} & {0.213}\tabularnewline
\bottomrule 
\multicolumn{7}{c}{{$\mu\in[1,5,10,50,100,150,200,250,500,750,1000]$}}\tabularnewline 
\end{tabular}{\scriptsize \par}}\hfill{}
\end{table}

\begin{table}
\caption{Selected weight parameter combinations parameter, which are stable
across folds, with training MAP. Left: grid tuning; Right: RankLib
(Figure \ref{fig:results-grid} shows results on test set).\label{tab:tuning-lambda}}

\noindent \centering{}{\scriptsize{}}%
\begin{tabular}{c@{\hskip 1em}llll@{\hskip 3em}llll}
\toprule 
{method} & {$\lambda_{u}$} & {$\lambda_{b}$} & {$\lambda_{w}$} & {MAP} & {$\lambda_{u}$} & {$\lambda_{b}$} & {$\lambda_{w}$} & {MAP}\tabularnewline
\midrule
{mrfSDM} & {0.85} & {0.15} & {0.05} & { 0.26} & {0.88} & {0.06} & {0.06} & {0.26}\tabularnewline
{genSDM} & {0.05} & {0.05} & {0.9} & { 0.21} & {0.32} & {0.45} & {0.24} & {0.26}\tabularnewline
{genNGram} & {0.35} & {0} & {0.65} & { 0.23} & {0.10} & {0.01} & {0.89} & {0.26}\tabularnewline
\bottomrule
\multicolumn{5}{r}{{\scriptsize{}$\lambda\in[0.0,0.05,0.10,\dots0.95,1.0]$}} & \multicolumn{4}{c}{coord ascent}\tabularnewline
\end{tabular}{\scriptsize \par}
\end{table}

\subsection{Empirical Results}

The results of the evaluation with standard error bars are presented
in Figure \ref{fig:results-grid} for the grid tuning experiment and
in Figure \ref{fig:results-ranklib} for the RankLib experiment. 

In the grid-tuning experiment it appears that the MRF-based SDM model is clearly better
than any of the other variants, including both generative models as
well as the MRF-variant with n-gram features. The second best method
is the query likelihood method. However, once $\boldsymbol{\lambda}$
is learned with coordinate ascent from RankLib, the difference disappears. This is concerning, because it may lead to the false belief of discriminative models being superior for this task.

The achieved performance of mrfSDM in both cases is consistent with
the results of the experiment conducted by Huston et al.~\cite{huston2013termdependencies}.

\paragraph{Generative models}

In all cases, weight parameters $\lambda$ and Dirichlet scale parameters
$\mu$ selected on the training folds, cf. Tables \ref{tab:tuning-mu}
and \ref{tab:tuning-lambda}, are stable across folds.

We observe that selected weight parameterization for the genNGram
model puts the highest weight on the windowed bigram model, omitting
the bigram model completely. In fact, among all four bigram language
models, the n-gram windowed bigram model, described in Section \ref{subsec:Generative-windowed-bigram-proce} achieves the highest retrieval
performance by itself (MAP 0.21, column $\mu_w$ in Table \ref{tab:genWindowedBigram}).

For the genSDM model, which is based on bag-of-bigrams, the weight parameters rather inconsistent across folds and training methods, suggesting that the model is unreliably when trained with cross validation.

\paragraph{Markov random fields}

In order to understand whether the success factor of the mrfSDM lies
in the log-linear optimization, or in the bag-of-bigram features,
we also integrate the n-gram based features discussed in Section \ref{sec:genNgram}
as features into the MRF-based SDM algorithm introduced by Metzler
et al.\ (discussed in Section \ref{sec:mrfSDM}). This approach is denoted
as mrfNGram in Figure \ref{fig:results-grid}. While the performance
is diminished when using grid-tuning, identical performance is achieved when parameters are estimated with RankLib (Figure \ref{fig:results-ranklib}).

\paragraph{Discussion}

We conclude that all four term-dependency methods are able to achieve the same
performance, no matter whether
a generative approach or a different bigram paradigm is chosen.
 We also do not observe any difference across levels
of difficulty (result omitted). This is not surprising given the similarities between
the models, as elaborated in this paper.

However, a crucial factor in this analysis is the use of a coordinate ascent
algorithm for selection of weight parameters. The coordinate ascent
algorithm was able to find subtle but stable weight combinations that
the grid tuning algorithm did not even inspect.

An important take-away is to not rely on grid tuning for evaluating discriminative model in comparison generative models, as it may falsely appear that the discriminative model achieves a significant performance improvement (compare mrfSDM versus genSDM in Figure \ref{fig:results-grid}), where actually this is only due to inabilities of fixed grid-searches to suitably explore the parameter space.

\section{Related Work}

\label{sec:Related-work}This work falls into the context of other
works that study different common axiomatic paradigms \cite{zhai2011axiomatic}  used in information retrieval
empirically and theoretically. Chen and Goodman \cite{chen1996smoothing}
studied different smoothing methods for language modeling, while Zhai
and Lafferty \cite{zhai2001smoothing} re-examine this question for
the document retrieval task. Finally, Smucker and Allan \cite{smucker2006smoothing}
concluded which characteristic of Dirichlet smoothing leads to its
superiority over Jelinek-Mercer smoothing.

Our focus is on the theoretical understanding of equivalences of different probabilistic models that consider sequential term dependencies, such as \cite{metzler2005sdm}. Our work is motivated to complement the empirical comparison
of Huston and Croft \cite{huston2013termdependencies,huston2014termdepedencies-appendix}.
Huston and Croft studied the performance of the sequential dependence
model and other widely used retrieval models with term dependencies
such as BM25-TP, as well as Terrier's pDFR-BiL2 and pDFR-PL2 with
an elaborate parameter tuning procedure with five fold cross validation.
The authors found that the sequential dependence model outperforms
all other evaluated method with the only exception being an extension,
the weighted sequential dependence model \cite{bendersky2010wsdm}.
The weighted sequential dependence model extends the feature space
for unigrams, bigrams, and windowed bigrams with additional features
derived from external sources such as Wikipedia titles, MSN query
logs, and Google n-grams.

\section{Conclusion}

In this work we take a closer look at the theoretical underpinning
of the sequential dependence model. The sequential dependence model
is derived as a Markov random field, where a common choice for potential
functions are log-linear models. We show that the only difference
between a generative bag-of-bigram model and the SDM model is that
one operates in log-space the other in the space of probabilities.
This is where the most important difference between SDM and generative
mixture of language models lies. 

We confirm empirically, that all four term-dependency models are capable
of achieving the same good retrieval performance. However, we observe
that grid tuning is not a sufficient algorithm for selecting the weight
parameter---however a simple coordinate ascent algorithm, such as
obtainable from the RankLib package finds optimal parameter settings. 
A shocking result is that for the purposes of comparing different models, 
tuning parameters on an equidistant grid may lead to the false belief that the MRF model is significantly better, where in fact, this is only due to the use of an insufficient parameter estimation algorithm.

This analysis of strongly related models that following the SDM model
in spirit, but are based on MRF, generative mixture models, and Jelinek-Mercer/interpolation
smoothing might appear overly theoretical. However, as many extensions
exist for the SDM model (e.g., including concepts or adding spam features)
as well as for generative models (e.g., relevance model (RM3), translation
models, or topic models), elaborating on theoretical connections and
pinpointing the crucial factors are important for bringing the two
research branches together. The result of this work is that, when
extending current retrieval models, both the generative and Markov
random field framework are equally promising.

\section*{Acknowledgements}
\small{This work was supported in part by the Center for Intelligent Information Retrieval. Any opinions, findings and conclusions or recommendations expressed in this material are those of the authors and do not necessarily reflect those of the sponsor.}

\bibliographystyle{plain}
\bibliography{theory-sdm}

\appendix 
\section{Approximations in Galago}

We noticed some approximations in Galago's implementation with respect
to the bigram and windowed bigram model which also affects the Dirichlet
smoothing component. For completeness we discuss
these approximations and their effects.

The denominator of both the model and the smoothing term provides
a normalizer reflecting counts of 'all possible cases'. In the unigram case, the counts of 'all possible cases' is the
document length $n_{\star,d}=|d|$ and for the smoothing component
the collection length $C=n_{\star,\star}=\sum_{d}|d|$.

In the bigram case, the number of all possible bigrams in a document
$n_{\left\{ \star,\star\right\} _{8},d}=|d|-1\approx|d|$ is approximated in
the implementation with the document length.
The approximation factors into the smoothing component $n_{\left(\star,\star\right),\star}=\sum_{d}\left(|d|-1\right)=C-\tilde{C}\approx C$
with $\tilde{C}$ denoting the number of documents in the collection.
For documents that are long on average, this is a reasonable approximation.

In the windowed-bigram case, all possible windowed bigrams
in a document $n_{\left\{ q_{i},q_{i+1}\right\} _{8},d}=\left(|d|-7\right)\cdot 28\approx|d|$. This is 
because the document has $|d|-7$ windows, each with 8 choose 2 cases. The approximation of off by a factor of 28. This also affects the smoothing component, $n_{\left\{ \star,\star\right\} _{8},\star}\approx\left(C-7\tilde{C}\right)\cdot 28\approx C$. However, when the smoothing parameter $\mu$ is tuned with relevance data, the constant factor of 28 is
absorbed by $\mu$.

\end{document}